\documentstyle[aps,prl,multicol,fancyheadings,rotate,epsfig,amsbsy,amssymb,amstex]{revtex}
\def\one{{\mathchoice {\rm 1\mskip-4mu l} {\rm 1\mskip-4mu l} {\rm
1\mskip-4.5mu l} {\rm 1\mskip-5mu l}}}
\def\bbbc{{\mathchoice {\setbox0=\hbox{$\displaystyle\rm C$}\hbox{\hbox
to0pt{\kern0.4\wd0\vrule height0.9\ht0\hss}\box0}}
{\setbox0=\hbox{$\textstyle\rm C$}\hbox{\hbox
to0pt{\kern0.4\wd0\vrule height0.9\ht0\hss}\box0}}
{\setbox0=\hbox{$\scriptstyle\rm C$}\hbox{\hbox
to0pt{\kern0.4\wd0\vrule height0.9\ht0\hss}\box0}}
{\setbox0=\hbox{$\scriptscriptstyle\rm C$}\hbox{\hbox
to0pt{\kern0.4\wd0\vrule height0.9\ht0\hss}\box0}}}}

\begin{document}
\title{
Unveiling Order behind Complexity: \\
Coexistence of Ferromagnetism and Bose-Einstein Condensation}
\author{C.D. Batista, G. Ortiz, and J.E. Gubernatis}
\address{Theoretical Division,
Los Alamos National Laboratory, Los Alamos, NM 87545}
\date{Received \today }
\maketitle

\begin{abstract}
We present an algebraic framework for identifying the order parameter
and the possible phases of quantum systems that is based on identifying
the local dimension $N$ of the quantum operators and using the $SU(N)$
group representing the generators of generalized spin-particle
mappings. 
We illustrate this for $N$=3 by presenting for any spatial dimension
the exact solution of the bilinear-biquadratic $S$=1 quantum Heisenberg
model at a high symmetry point. Through this solution we
rigorously show that itinerant ferromagnetism and Bose-Einstein
condensation may coexist.
\end{abstract}


\vspace*{-0.4cm}

\begin{multicols}{2}

\columnseprule 0pt

\narrowtext
\vspace*{-0.5cm}

{\it Introduction.} One of the fascinating things about matter is the
different phases it exhibits. One way or another, identifying these
phases and the transitions between them occupies the interests and
activities of most condensed-matter physicists, physical chemists, and
field theorists. What is typically used as a working principle is
Landau's postulate of an order parameter (OP). While one generally
knows what to do if the OP is known, Landau's postulate gives no
procedure for finding it. This problem is particularly complex
if the symmetries of the system are hidden and ``veil'' the nature of
the order. For example, it took a while to recognize that the staggered
magnetization was the OP for anti-ferromagnetism.

In this Letter, we present a simple algebraic framework
for identifying OPs and the possible zero temperature phases (ground
states (GSs)) of quantum systems of spins, fermions, and bosons.
Central to this framework is identifying the local dimension $N$ of the
quantum operators, which sets the order of the largest Lie group that
can be used to describe those systems. The OP is constructed from this
hierarchical group, and the embedded subgroups are identified to
establish a hierarchical classification of the  possible broken
symmetry phases. 

To place this framework in more familiar terms, we recall the
Jordan-Wigner \cite{jw} and Matsubara-Matsuda \cite{matsu}
spin-particle transformations that map $S=\frac{1}{2}$ spins into
spinless fermions and into hard-core (HC) bosons. In the latter case,
this mapping demonstrated that ferromagnetism in a localized spin model
was isomorphic to Bose-Einstein (BE) condensation in a HC boson model.
We note that these two venerable transformations are special cases of
the generalized transformations we recently introduced
\cite{ours1,Note0} and in fact they should be viewed as the same
transformation: Their local dimension is $N$=2, corresponding to the up
and down states of a spin or the occupied and unoccupied states of a
spinless fermion or a HC boson. The  hierarchical group is $SU(2)$, and
the broken symmetry states break either this symmetry in the
ferromagnetic (FM) case or the $U(1)$ symmetry of its subgroup in the
BE condensation case.

The key to the general framework we are proposing is the existence of a
general set of $SU(N)$ spin-particle transformations \cite{ours1,ours2},
fundamental for devising a general theoretical scheme to understand the
order hidden in complex behavior. More simply put, through the
generalized spin-particle transformations, features that are subtle and
hard to identify in one representation (hidden symmetries) can become
prominent and easy to analyze in another (explicit symmetries).
Indeed, these mappings {\it connect} seemingly unrelated physical
phenomena establishing equivalence relations among them. In the
hierarchical group, all elements of the operator basis are symmetry
generators \cite{Note00}. This allows one to study the coexistence and
competition of phases, like magnetism and superconductivity, with the
corresponding OPs derived from the subgroup generators embedded in the
largest global symmetry group of the problem.

The $N$=2 case is too simple to illustrate all these points. We can,
however, illustrate them by moving up to $N$=3 and be very concrete by
studying a $S$=1 bilinear-biquadratic quantum spin Hamiltonian. In
fact, we will present the exact solution to this problem at a point of
high symmetry in any spatial dimension. Using our generalized
spin-particle transformations, we show this model happens to map onto a
lattice gas model of interacting particles with an internal quantum
number (two-flavored HC bosons). The exact solution
transparently exhibits two simultaneously broken continuous symmetries
associated to the formation of a polarized BE condensate. In this way,
we rigorously show that these two quantum orderings can coexist
\cite{suto}. This novel result emerges naturally by identifying the
underlying highest rank algebra ($su(3)$), providing
the framework to determine the possible broken symmetries and
understand the role of symmetry-reducing perturbations. Using it, we
will construct the complex phase diagram in a neighborhood of the
model's high symmetry point.


{\it Ferromagnetism and BE Condensation.} The Hamiltonian we studied is
\cite{papa}
\begin{equation} \hspace*{-0.9cm}
H_\theta = J\sqrt{2} \sum_{\langle {\bf i},{\bf j} \rangle} \left [
\cos \theta \ {\bf S}_{\bf i} \cdot {\bf S}_{{\bf j}} + \sin \theta
\left ( {\bf S}_{\bf i} \cdot {\bf S}_{{\bf j}} \right )^2 \right ] ,
\label{spin1}
\end{equation}
where $J<0$ and summation is over bonds $\langle {\bf i},{\bf j}
\rangle$ of a regular $d$-dimensional lattice with $N_s$ sites. A
spin $S$=1 operator ${\bf S}_{\bf i}$ is associated with lattice
site ${\bf i}$ and locally satisfies the $su(2)$ Lie algebra.

The case $H_1=H_{\theta=\frac{\pi}{4}}$ can be conveniently
written in a ($s=\frac{1}{2}$) HC boson representation 
($\bar{n}_{{\bf j} \sigma}\bar{n}_{{\bf j}
\sigma'}= \delta_{\sigma \sigma'} \bar{n}_{{\bf j} \sigma}$)
\cite{Note1}
\begin{eqnarray}
\frac{S^+_{\bf j}}{\sqrt{2}} = \bar{b}^\dagger_{{\bf j} \uparrow} +
\bar{b}^{\;}_{{\bf j} \downarrow} \ , \
\frac{S^-_{\bf j}}{\sqrt{2}} = \bar{b}^{\;}_{{\bf j} \uparrow} +
\bar{b}^\dagger_{{\bf j} \downarrow} \ , \
S^z_{\bf j} = \bar{n}_{{\bf j} \uparrow} - \bar{n}_{{\bf j}
\downarrow} \ ,
\label{map}
\end{eqnarray}
as an extended $t$-$J$ like Hamiltonian,
\begin{eqnarray}
H_1&=&J \! \sum_{\langle {\bf i},{\bf j} \rangle,\sigma} \left (
{\bar{b}}^\dagger_{{\bf i} \sigma} {\bar{b}}^{\;}_{{\bf j} \sigma}
+ {\rm H.c.} \right ) + 2 J \sum_{\langle {\bf i},{\bf j} \rangle}
{\bf s}_{\bf i} \cdot  {\bf s}_{\bf j} \nonumber \\
&+& 2 J \sum_{\langle {\bf i},{\bf j} \rangle} \left ( 1 -
\frac{{\bar{n}}_{\bf i}+{\bar{n}}_{\bf j}}{2} +
\frac{3}{4} {\bar{n}}_{\bf i} {\bar{n}}_{\bf j}  \right ) \ ,
\label{hamilt}
\end{eqnarray}
with ${\bf s}_{\bf j}=\frac{1}{2} {\bar {b}}^\dagger_{{\bf j} \mu}
\boldsymbol{\sigma}_{\mu \nu} {\bar {b}}^{\;}_{{\bf j} \nu}$ a
$s=\frac{1}{2}$ operator ($\boldsymbol{\sigma}$ denoting Pauli
matrices), and ${\bar {n}}_{\bf j}={\bar {b}}^\dagger_{{\bf j}
\uparrow} {\bar {b}}^{\;}_{{\bf j} \uparrow}+ {\bar
{b}}^\dagger_{{\bf j} \downarrow} {\bar {b}}^{\;}_{{\bf j}
\downarrow}$. This last form in turn can be rewritten as
\begin{equation}
H_1=2 J \sum_{\langle {\bf i},{\bf j} \rangle} P_s({\bf i},{\bf j})
\ ,
\end{equation}
where $P_s({\bf i},{\bf j})=P^2_s({\bf i},{\bf j})$ is the projector
onto the symmetric subspace ($S$=0,2) corresponding to the bond
$\langle {\bf i},{\bf j} \rangle$ which indicates that if one finds a
state that is symmetric under the permutation of nearest neighbors
${\bf r}_{\bf i}$ and ${\bf r}_{\bf j}$, then that state is the GS.


For a system of ${\cal N}$ HC bosons the GS is
\begin{equation}\label{GS}
| \Psi_0({\cal N},S_z) \rangle = (\tilde{b}^{\dagger}_{{\bf 0}
\uparrow})^{{\cal N}_{\uparrow}} (\tilde{b}^{\dagger}_{{\bf
0}\downarrow})^{{\cal N}_{\downarrow}} | 0 \rangle \ ,
\end{equation}
(${\cal N}= {\cal N}_{\uparrow}+{\cal N}_{\downarrow} \leq N_s$) with
an energy $E_0=J N_s {\sf z}$ (${\sf z}$ is the coordination of the
lattice) and a total $S_z = \frac{{\cal N}_{\uparrow}-{\cal
N}_{\downarrow}}{2}$. The operator $\tilde{b}^{\dagger}_{{\bf
0}\sigma}$ is the ${\bf k}={\bf 0}$ component of 
$\bar{b}^\dagger_{{\bf j}\sigma}$, i.e., $\tilde{b}^\dagger_{{\bf k}
\sigma}= \frac{1}{\sqrt{N_s}} \sum_{\bf j} \exp[i{\bf k}\cdot {\bf
r}_{\bf j}] \ {\bar{b}}^\dagger_{{\bf j} \sigma}$. The quasihole and
quasiparticle excited states are
\begin{eqnarray}
                \begin{cases}
| \Psi^h_{\bf k}({\cal N},S_z) \rangle = \tilde{b}^{\;}_{{\bf k}
\sigma} | \Psi_0({\cal N},S_z)\rangle &
        \text{quasihole}, \\
| \Psi^p_{\bf k}({\cal N},S_z) \rangle = \tilde{b}^{\dagger}_{{\bf
k} \sigma} | \Psi_0({\cal N},S_z)\rangle &
        \text{quasiparticle} ,
        \end{cases}
\end{eqnarray}
with the excitation energy of each being $\omega_{\bf k} = J{\sf z} 
(\frac{1}{\sf z} \sum_{\nu} e^{i {\bf k} \cdot {\bf e}_\nu}-1)$ where the
sum runs over the vectors ${\bf e}_\nu$ which connect a given site to
its ${\sf z}$ nearest neighbors. In the  $|{\bf k}| \rightarrow 0$
limit, $\omega_{\bf k}\rightarrow 0$.

Clearly the GS in Eq.~\ref{GS} is a FM BE condensate with any partial
spin polarization, and the form of the result is {\it independent of
the spatial dimensionality of the lattice}. We note that different
values of $S_z$ correspond to the different orientations of the
magnetization ${\cal M}$ associated to the broken $SU(2)$ spin
rotational symmetry of the GS. We also note that the degeneracy of
states with different number of particles ${\cal N}$ indicates a broken
$U(1)$ charge symmetry (conservation of the number of particles)
associated to the BE condensate \cite{Note2}. A signature of BE
condensation is the existence of off-diagonal long-range order (ODLRO) 
in the correlation function $\Phi_{\sigma \sigma'}({\bf ij})=\langle 
\bar{{b}}^\dagger_{{\bf i} \sigma} \bar{{b}}^{\;}_{{\bf j}\sigma'} 
\rangle$. When ${\cal N}_\uparrow$ and ${\cal N}_\downarrow$ are both
of order $N_s$, there are two eigenvectors with eigenvalues of order
$N_s$ and the condensate is thus a mixture.  

\begin{figure}[htb]
\epsfverbosetrue
\epsfxsize=8cm
\centerline{\epsfbox{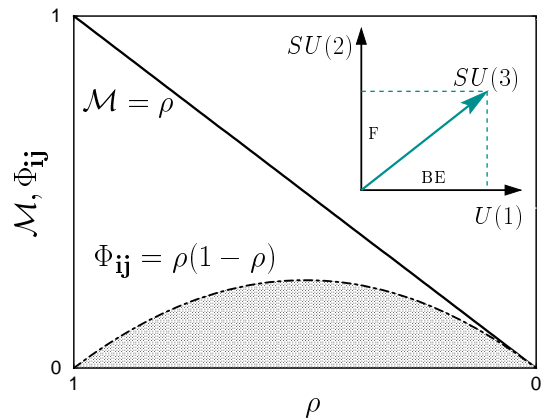}}
\vspace*{-0.0cm} 
\caption{Coexistence of ferromagnetism and Bose-Einstein condensation.
The correlation function $\Phi_{\uparrow \uparrow}({\bf ij})=\Phi_{\bf
ij}$ is expressed in the thermodynamic limit.  
The inset schematically displays the $SU(3)$ order parameter living in
an eight-dimensional space with projections onto the $SU(2)$ (F:
ferromagnet) and $U(1)$ (BE: Bose-Einstein condensate) axes.}
\label{fig1}
\end{figure}

We can easily compute the magnetization ${\cal M}$ and phase coherence
of these various (non-normalized) degenerate GSs for a given density
$\rho=\frac{{\cal N}}{N_s}$.  For example, in the fully polarized case,
${\cal N}={\cal N}_\uparrow$, ${\cal M}=\langle S^z_{\bf j} \rangle =
\rho$, and the ODLRO $\Phi_{\uparrow \uparrow}({\bf ij})= \frac{\rho
(1-\rho)}{1-\epsilon}$ (${\bf r}_{\bf i} \neq {\bf r}_{\bf j}$), where
$\epsilon=1/N_s$.  Similarly, the two-particle correlation function
$\langle \Delta^\dagger_{\bf i} \Delta^{\;}_{\bf j} \rangle =
\Phi_{\uparrow \uparrow}({\bf ij})
\frac{(\rho-\epsilon)(1-\rho-\epsilon)}{(1-2\epsilon) (1-3\epsilon)}$,
where $\Delta^\dagger_{\bf i}={\bar{b}}^\dagger_{{\bf i}
\uparrow}{\bar{b}}^\dagger_{{\bf i}+\boldsymbol{\delta} \uparrow}$
\cite{Note3}. Therefore the exact GS has two spontaneously broken
continuous symmetries (see Fig.~\ref{fig1}).

\vspace*{-.6cm}

The exact solution defines the features of the phase diagram that our
proposed framework must qualitatively admit. We will see below that
both OPs (magnetization and phase), as promised, are embedded in an
$SU(3)$ OP. We remark that the phase coexistence in the boson
representation maps back to a $S$=1 FM phase coexisting with
another spin phase. To see this consider the state $| \Psi_0({\cal
N},0) \rangle$ for which  $\langle S^z_{\bf i} \rangle =  \langle 
S^x_{\bf i} \rangle =  \langle  S^y_{\bf i} \rangle = 0$, which implies
that it is a singlet state in the $S$=1 representation. We will show
below that this other phase has a pure spin-nematic ordering.

What can be said about the fermionic extended $t$-$J$ Hamiltonian 
(i.e., Eq. \ref{hamilt} with HC bosonic operators replaced by
``constrained'' fermions)? Obviously, the GS energies of both
fermionic and bosonic Hamiltonians coincide at $\rho=0,1$. For
$0<\rho<1$, one can show that $E_0({\rm fermion}) > E_0({\rm boson})=
JN_s{\sf z}$, indicating that the fermionic system phase segregates
\cite{ours2}.

{\it The Framework Illustrated.} In Ref. \cite{ours2}, we generalized
the spin-particle transformations to spins satisfying the $SU(N)$ Lie
bracket relations. In particular, the fundamental (``quark'')
representations of $SU(N)$ are mapped onto an algebra of HC bosons
\cite{ours2} with $N_f=N-1$ flavors
\begin{eqnarray}
{\cal S}^{\alpha\beta}({\bf j})&=&{\bar{b}^{\dagger}_{{\bf j}\alpha}
\bar{b}^{\;}_{{\bf j}\beta} - \frac{\delta_{\alpha\beta}}{N}}  \ ,
\nonumber \\ {\cal S}^{\alpha 0}({\bf j})&=&\bar{b}^{\dagger}_{{\bf
j}\alpha} \; \ , \; {\cal S}^{0\beta}({\bf j})=\bar{b}^{\;}_{{\bf
j}\beta} \ , \nonumber \\ {\cal S}^{0 0}({\bf
j})&=&\frac{N_f}{N}-\sum^{N_f}_{\alpha=1} \bar{b}^{\dagger}_{{\bf
j}\alpha} \bar{b}^{\;}_{{\bf j}\alpha}= -\sum^{N_f}_{\alpha=1} {\cal
S}^{\alpha\alpha}({\bf j}) \ ,
\end{eqnarray}
where $1\leq \alpha,\beta \leq N_f$ run over the set of bosonic
flavors, ${\bf j}$ is the site index, and $\bar{b}^{\dagger}_{{\bf
j}\alpha}= {b}^{\dagger}_{{\bf j}\alpha} \prod_{\beta} (1-n_{{\bf
j}\beta})$. $S^{\mu \nu}({\bf j})$ ($0\leq \mu,\nu \leq N_f$) are the
components of the $SU(N)$ spin ($N^2-1$ linearly independent
components). These components are generators of an $su(N)$ Lie algebra
with commutation relations $[{\cal S}^{\mu \mu'}({\bf j}),{\cal S}^{\nu
\nu'}({\bf j})]= \delta_{\mu' \nu} {\cal S}^{\mu \nu'}({\bf
j})-\delta_{\mu \nu'} {\cal S}^{\nu \mu'}({\bf j})$. In particular, for
$N$=3 \cite{cartan}
\begin{equation}
{\cal S}({\bf j})= \begin{pmatrix} \frac{2}{3} - \bar{n}_{{\bf j}}
&\bar{b}^{\;}_{{\bf j} \uparrow} & \bar{b}^{\;}_{{\bf j} \downarrow} \\
\bar{b}^\dagger_{{\bf j} \uparrow}&\bar{n}_{{\bf j} \uparrow} -\frac{1}{3}&
\bar{b}^\dagger_{{\bf j} \uparrow} \bar{b}^{\;}_{{\bf j} \downarrow} \\
\bar{b}^\dagger_{{\bf j} \downarrow}&\bar{b}^\dagger_{{\bf j} \downarrow}
\bar{b}^{\;}_{{\bf j} \uparrow}&\bar{n}_{{\bf j} \downarrow}-\frac{1}{3}
\end{pmatrix} \ ,
\label{spinsu3}
\end{equation}
and up to an irrelevant constant we can rewrite $H_\theta$ as
\begin{eqnarray}
H_\theta&=& J \sqrt{2}\sum_{\langle {\bf i},{\bf j} \rangle}
\left [\cos \theta \ {\cal S}^{\mu \nu}({\bf i}) {\cal S}^{\nu
\mu}({\bf j})
\right . \nonumber \\
&+& \left . (\sin \theta - \cos \theta) \ {\cal S}^{\mu \nu}({\bf
i}) \tilde{\cal S}^{\nu \mu}({\bf j}) \right ]
\end{eqnarray}
where repeated Greek superscripts are summed and
\begin{equation}
\tilde{\cal S}({\bf j})= \begin{pmatrix} \frac{2}{3} - \bar{n}_{{\bf j}}
&-\bar{b}^{\dagger}_{{\bf j} \downarrow} & -\bar{b}^{\dagger}_{{\bf j} \uparrow}
\\ -\bar{b}^{\;}_{{\bf j} \downarrow}&\bar{n}_{{\bf j} \downarrow}
-\frac{1}{3}& \bar{b}^\dagger_{{\bf j} \uparrow} \bar{b}^{\;}_{{\bf j}
\downarrow} \\ -\bar{b}^{\;}_{{\bf j} \uparrow}&\bar{b}^\dagger_{{\bf
j} \downarrow} \bar{b}^{\;}_{{\bf j} \uparrow}&\bar{n}_{{\bf j}
\uparrow}-\frac{1}{3}
\end{pmatrix}
\label{spinsu}
\end{equation}
generates the conjugate representation on sublattice ${\bf j}$
\cite{Note4}. 
This expression for $H_\theta$ illustrates the very important result
that any nonlinear interaction in the original representation is simply
a bilinear term in the new representation when mapped onto the highest
rank algebra \cite{ours2}. In particular, there are certain special
points in parameter space where $H_\theta$ is highly symmetric.
For example, for $\theta = \frac{\pi}{4}$ and $\frac{5\pi}{4}$,
$H_\theta$ is explicitly invariant under uniform $SU(3)$
transformations on the spins \cite{sutherland}, while for $\theta=\pm
\frac{\pi}{2}$, $H_\theta$ is explicitly invariant under staggered
conjugate rotations of the two sublattices \cite{affleck}. These
symmetries are hard to identify in the original spin representation but
are evident in the HC bosonic representation.

The case where $\theta=\frac{\pi}{4}$ corresponds to the FM $SU(3)$
Heisenberg Hamiltonian, and therefore the GS is the state with maximum
total $SU(3)$ spin ${\cal S}$. The OP associated with this broken
symmetry is the total $SU(3)$ magnetization ${\cal S}^{\mu \nu }=
\frac{1}{N_s} \sum_{\bf j} {\cal S}^{\mu \nu }({\bf j})$ which has
eight independent components. When $\langle {\cal S}^{\mu \nu } \rangle
\neq 0$, the system orders, and the coexistence of a FM phase and a BE
condensation becomes more evident: In the bosonic language both OPs
correspond to different components of the $SU(3)$ OP (see Eq.
\ref{spinsu3} and Fig.~\ref{fig1}):
$\bar{b}^{\dagger}_{\uparrow}={\cal S}^{0 1}$,
$\bar{b}^{\dagger}_{\downarrow}={\cal S}^{0 2}$,
$\bar{b}^{\;}_{\uparrow}={\cal S}^{1 0}$, and
$\bar{b}^{\;}_{\downarrow}={\cal S}^{2 0}$, are the components of
the OPs for the $\uparrow$ and $\downarrow$ BE condensates, while
$s^z=\frac {1}{2}({\cal S}^{1 1}-{\cal S}^{2 2})$,
$s^x=\frac{1}{2}({\cal S}^{1 2}+{\cal S}^{2 1})$, and
$s^y=\frac{1}{2i}({\cal S}^{1 2}-{\cal S}^{2 1})$, are the
components of the FM OPs for the bosons. The ability of
$\langle\bar{n}_{\bf j} \rangle=\frac{2}{3}-\langle {\cal S}^{0 0}({\bf
j})\rangle$ to take any value during $SU(3)$ rotations of the GSs is
another manifestation of the BE condensation.

We will now work to answer the question: What is the OP of the model in
the original $S$=1 language (Eq. \ref{spin1})? We start by writing the
relation between the components of ${\cal S}^{\mu \nu}$ and the $S$=1
generators of $SU(2)$. From Eqs. \ref{map} and \ref{spinsu3} we have
\begin{eqnarray}
S^x&=&\frac{1}{\sqrt{2}} ({\cal S}^{0 1}+{\cal S}^{2 0}+ {\cal S}^{0
2}+{\cal S}^{1 0}) \ , \nonumber \\
S^y&=&\frac{1}{\sqrt{2}i} ({\cal S}^{0 1}+{\cal S}^{2 0}-{\cal S}^{0
2}-{\cal S}^{1 0}) \ , \nonumber \\
S^z&=&{\cal S}^{1 1}-{\cal S}^{2 2} \ ,
\nonumber \\
(S^x)^2-\frac{2}{3}&=&\frac{1}{2}({\cal S}^{1 2}+{\cal S}^{2 1}+{\cal
S}^{0 0}) \ , \ (S^z)^2-\frac{2}{3}= -{\cal S}^{0 0}  \ , \nonumber
\\
\left \{ S^x,S^y \right \}&=& i ({\cal S}^{2 1}-{\cal S}^{1 2}) \ ,
\nonumber \\
\left \{ S^x,S^z \right \}&=&\frac{1}{\sqrt{2}} ({\cal S}^{0 1}-{\cal
S}^{2 0}-{\cal S}^{0 2}+{\cal S}^{1 0}) \ , \nonumber \\
\left \{ S^y,S^z \right \}&=&\frac{1}{\sqrt{2}i} ({\cal S}^{0 1}-{\cal
S}^{2 0}+{\cal S}^{0 2}-{\cal S}^{1 0}) \ .
\end{eqnarray}
The first three operators correspond to the $S$=1 FM OP, while the
second five are the components of the spin-nematic OP (components of
the bilinear symmetric traceless tensor). The traceless condition of
${\cal S}^{\mu \nu}$ implies that $(S^y)^2-\frac{2}{3} =
-((S^z)^2-\frac{2}{3}) - ((S^x)^2-\frac{2}{3})$.  This classification
is independent of the parity and time-reversal properties of
$\Lambda^{ab}(t)= S^a(t)S^b(0)$ ($a,b=x,y,z$) \cite{sasha}. We see that
by rotating the $SU(3)$ FM OP, we find a FM or a spin-nematic GS for
the $S$=1 Hamiltonian. Moreover, we also see that the existence of a
spin-nematic quantum phase is directly related to the BE condensation
of the two-flavored HC bosons. Concomitantly, we have rigorously
established the existence of a spin-nematic phase, an issue that had
been outstanding in the literature \cite{solyom}.

{\it Unveiling Order and Establishing Complexity.} With the phases of
$H_\theta$ established at the point $\theta=\frac{\pi}{4}$ we will
now detail the nature of the phases near it as revealed by adding a
symmetry breaking field to the model. A vast array of rich behavior is
discovered, and we now highlight some of the more interesting results.

Coupling our two-flavored bosonic system $H_1$ to a magnetic field
${\bf B}$ through a Zeeman term removes the degenerancy of the FM GS
associated to the $SU(2)$ symmetry. The resulting GS is a BE condensate
polarized along the magnetic field direction (chosen to be the
longitudinal $z$ direction). In fact, in the limit $|{\bf B}|
\rightarrow \infty$, the resulting Hamiltonian is $H_{t\!-\!V}= t\!
\sum_{\langle {\bf i},{\bf j} \rangle} ({\bar{b}}^\dagger_{{\bf i}
\uparrow} {\bar{b}}^{\;}_{{\bf j} \uparrow} + \bar{b}^{\dagger}_{{\bf
j} \uparrow}\bar{b}^{\;}_{{\bf i} \uparrow}) + V\! \sum_{\langle {\bf
i},{\bf j}\rangle} ({\bar{n}}_{{\bf i}\uparrow}-\frac{1}{2})
({\bar{n}}_{{\bf j}\uparrow}-\frac{1}{2})$ and represents a gas of HC
bosons with hopping $t=J$ and nearest-neighbor density-density
interaction $V=2J$.

By means of the Matsubara-Matsuda transformation \cite{matsu}, we can
map $H_{t-V}$ onto the FM $S=\frac{1}{2}$ Heisenberg model which has an
exact solution. Therefore, ${\bf B}$ removes the $SU(3)$ invariance of
$H_1$ and leaves an $SU(2)$ invariance associated to the charge
degrees of freedom. In this way we recover the well-known connection
between a FM $S =\frac{1}{2}$ state in the transverse direction and a
BE condensate of spinless HC bosons. Because of the $SU(2)$ invariance
of the Heisenberg model, the FM states polarized along $z$ are also
GSs, so if we infinitesimally increase the attractive interaction
between bosons, the model phase segregates (Ising-like ferromagnetism).

For $V=0$, $H_{t\!-\!V}$ reduces to an XY model whose GS is again
a BE condensate (i.e., has off-diagonal long range order
\cite{Lieb}). Regardless of the dimension of the system, the GS of
$H_{t\!-\!V}$ is a BE condensate for $-2|t|\le V\le 0$ and phase
segregates when $V<-2|t|$. In fact, for a very small repulsive
interaction  the GS is still a BE condensate \cite{Lieb}. In a
similar fashion $H_1$ is at the point separating the BE
condensate from phase segregation, i.e., if we make the two-body
nearest-neighbor interaction infinitesimally more attractive, there
is phase segregation.

A possible realization of the single-flavor bosonic system
$H_{t\!-\!V}$ is the attractive $U$ Hubbard model in the large $|U|$
limit \cite{Nozieres2} with $t=-2\tau^2/|U|$ and $V_1=4\tau^2/|U|$
($\tau$ is the hopping integral of the original Hubbard model). It is
clear that if we add a nearest-neighbor attractive interaction
$v=-2\tau^2/|U|$ to the original Hubbard model, we can exactly compute
the superconducting GS and quasiparticle excitations in any spatial
dimension and for any concentration of particles since $V=V_1+4v=2t$.
In the case where $v=-\tau^2/|U|$ and $V=0$, the Hamiltonian
is an XY model whose GS displays long-range order \cite{Lieb}.


{\it Conclusions.} 
We presented an algebraic framework aimed at uncovering the order
behind the potential multiplicity of complex phases in interacting
quantum systems \cite{bohm}. In this framework the local Hilbert space
of dimension $N$ admits a convenient hierarchical representation in
terms of the generators of an $SU(N)$ group \cite{ours2}. We
illustrated this for $N$=3 by first presenting the exact solution of a
non-trivial model of interacting quantum spins. A key point is that
from this hierarchical group ($SU(3)$ in the present case), one can
identify the embedded subgroups (an invariant $SU(2)$ for spin and an
$SU(2)$ for charge and two others connecting them), thereby
establishing a hierarchical classification of order parameters.

Through the spin model, we rigorously demonstrated the possible
coexistence of ferromagnetism and BE condensation of $s=\frac{1}{2}$
hard-core bosons. This condensate was shown to be related to a quantum
spin-nematic ordering in the original spin model. These results may
shed light on the physics associated with the recent surprising
experimental observation of the coexistence of superconductivity and
ferromagnetism in UGe$_2$ \cite{experim}.

We close with the observation that a fundamental relation between
magnetism and BE condensation has emerged as a consequence of our
spin-particle transformations: If a solution of a particular localized
quantum spin $S$ model displays long-range order in the transverse
direction ($\lim_{|{\bf r}_{\bf i}-{\bf r}_{\bf j}| \rightarrow \infty}
\langle S^a_{\bf i}S^a_{\bf j} \rangle \neq 0$, with $a=x,y$), then the
corresponding itinerant bosonic model displays long-range phase
coherence (i.e., a BE condensation). This result is general and
independent of the sign of the exchange couplings in the spin model. 

%

We thank D. Pines for a useful discussion. This work was sponsored by
the US DOE under contract W-7405-ENG-36. Part of it was performed while
G. O. was a visitor at the Institute for Theoretical Physics (UCSB)
which is supported by the grant No. PHY99-07949.

\end{multicols}


\begin{references}
\bibitem{cartan}
In the Cartan treatment of Lie algebras \cite{cornwell}, the Cartan
subalgebra of $SU(3)$ has rank 2 and our generators are ${\cal
S}^{11}=$H$_1$ and ${\cal S}^{22}=$H$_2$ while the remaining
off-diagonal elements of the matrix Eq.~\ref{spinsu3} represent the
step operators E$_i$ ($i=1,2,\cdots,6$). These eight generators form
the Cartan-Weyl basis of the Lie algebra $su(3)$.
\end{references}
\end{document}